\documentclass[conference]{IEEEtran}
\usepackage{authblk}
\IEEEoverridecommandlockouts
\usepackage{cite}
\usepackage{amsmath,amssymb,amsfonts}
\usepackage{graphicx}
\usepackage{xcolor}

\usepackage{enumitem}
\usepackage[table]{xcolor}
\usepackage{multirow}
\definecolor{Gray}{gray}{0.9} % Light gray for row 
\definecolor{LightBlue}{RGB}{173, 216, 230} % Light blue
\usepackage{textcomp}
\usepackage{xcolor}
\usepackage{graphicx} % For including graphics if needed
\usepackage{booktabs} % For enhanced table layout with \toprule, \midrule, \bottomrule
\usepackage{url}
\usepackage{amsmath}
\usepackage{amssymb}
\usepackage{algorithm}
\usepackage{algorithmic}
\usepackage[dvipsnames]{xcolor}
\usepackage{subcaption}
\usepackage{makecell}
\usepackage{pdfpages}
\usepackage{array} % For better array handling
\usepackage[most]{tcolorbox}
\usepackage[table,xcdraw]{xcolor} % For coloring rows and cells
\usepackage{caption} % For managing captions, optional 
\def\BibTeX{{\rm B\kern-.05em{\sc i\kern-.025em b}\kern-.08em
    T\kern-.1667em\lower.7ex\hbox{E}\kern-.125emX}}

\begin{document}

\date{16 Sept 2025}

\title{AQUA-LLM: Evaluating Accuracy, Quantization, and Adversarial Robustness Trade-offs in LLMs for Cybersecurity Question Answering}

%\author{}
\author{Onat Gungor$^{*}$}
\author{Roshan Sood$^{*}$}
\author{Harold Wang}
\author{Tajana Rosing}

\affil{Department of Computer Science and Engineering, University of California, San Diego}
\affil{\{ogungor, rosood, haw079, tajana\}@ucsd.edu}

\maketitle

\begingroup
\renewcommand\thefootnote{\textasteriskcentered}
\footnotetext{Both authors contributed equally to this research.}
\endgroup

\pagestyle{plain}
\pagenumbering{gobble}
\newcommand{\norm}[1]{\left\lVert#1\right\rVert}
\newcommand{\Design}[0]{AQUA-LLM}

\begin{abstract}
Large Language Models (LLMs) have recently demonstrated strong potential for cybersecurity question answering (QA), supporting decision-making in real-time threat detection and response workflows. However, their substantial computational demands pose significant challenges for deployment on resource-constrained edge devices. Quantization, a widely adopted model compression technique, can alleviate these constraints. Nevertheless, quantization may degrade model accuracy and increase susceptibility to adversarial attacks. Fine-tuning offers a potential means to mitigate these limitations, but its effectiveness when combined with quantization remains insufficiently explored. Hence, it is essential to understand the trade-offs among accuracy, efficiency, and robustness. We propose AQUA-LLM, an evaluation framework designed to benchmark several state-of-the-art small LLMs under four distinct configurations: base, quantized-only, fine-tuned, and fine-tuned combined with quantization, specifically for cybersecurity QA. Our results demonstrate that quantization alone yields the lowest accuracy and robustness despite improving efficiency. In contrast, combining quantization with fine-tuning enhances both LLM robustness and predictive performance, achieving an optimal balance of accuracy, robustness, and efficiency. These findings highlight the critical need for quantization-aware, robustness-preserving fine-tuning methodologies to enable the robust and efficient deployment of LLMs for cybersecurity QA.    
\end{abstract}

\section{Introduction}
The evolving complexity of cyber threats poses significant challenges for cybersecurity professionals \cite{dekker2024threat}. In 2024, large enterprises spent an average of \$14.6 million annually on Security Operations Centers, with around 80\% allocated to labor costs \cite{kpmg2024cybersecurity}, highlighting the urgent need for scalable, intelligent solutions to ease operational and cognitive burdens. Large Language Models (LLMs) present a promising avenue by enhancing intelligence, accessibility, and adaptability in cybersecurity workflows \cite{tian2025exploring}. Leveraging LLMs enables systems to analyze complex security data, deliver actionable insights, and support decision-making, improving operational efficiency and effectiveness \cite{ferrag2024generative}. Their applications span automated incident response, malware detection, vulnerability assessment, and numerous other cybersecurity tasks \cite{xu2024large}. Among these, cybersecurity question answering (QA) is a key application that automates precise responses to security queries, enabling rapid threat identification and remediation \cite{agrawal2024cyberq}. LLMs are particularly valuable for this task due to their capacity to understand and generate contextually relevant, human-like responses \cite{rajapaksha2024rag}.

%%%%%%%%%%%%%%%%%%%%%%%%%%%%%%%%%%%%%%%%%%%%%%%%%%%
%\begin{tcolorbox}[colback=gray!5, colframe=black, title=\textbf{Cybersecurity %QA Example}, sharp corners=all]
%\textbf{Question:} What is the appropriate mitigation strategy for a Distributed Denial-of-Service (DDoS) attack detected in network traffic?

%\textbf{Answer:} Mitigation strategies include rate limiting, traffic filtering, deploying a Web Application Firewall, and using cloud-based DDoS protection services.
%\end{tcolorbox}
%%%%%%%%%%%%%%%%%%%%%%%%%%%%%%%%%%%%%%%%%%%%%%%%%%%

Despite their promise, LLM-based cybersecurity QA systems face significant deployment challenges in real-time, resource-constrained environments. The need for fast, context-aware responses is especially critical at the network edge, where data privacy and low latency are essential. For instance, edge devices in power grids can use LLMs to interpret intrusion alerts and support operators with local, actionable insights. However, state-of-the-art approaches \cite{rajapaksha2024rag, tihanyi2024cybermetric} rely on large proprietary models such as GPT, which demand extensive storage and computational resources. Although solutions such as edge-cloud collaboration and hardware acceleration have been proposed, model optimization is among the most widely applicable. In particular, quantization reduces memory and compute overhead by lowering numerical precision, though often at the cost of accuracy and robustness \cite{shen2024agile, zhang2024edge}.

Quantization facilitates the deployment of LLMs on edge devices by reducing resource requirements but often leads to decreased predictive accuracy~\cite{husom2025sustainable}. Concurrently, quantization may compromise model robustness, increasing vulnerability to adversarial manipulation~\cite{li2024investigating, kumar2024fine}. This poses significant risks in safety-critical cybersecurity question answering applications, where failures can result in exposure of sensitive information or generation of misleading responses. Fine-tuning has emerged as a promising approach to mitigate these issues. Prior work has leveraged fine-tuning to recover the predictive performance of quantized LLMs~\cite{wang2024taming, zhou2025quzo}, and it has also been shown to enhance adversarial robustness in standard LLMs~\cite{o2023adversarial, liu2024adversarial}. However, the impact of fine-tuning on the robustness of quantized models remains underexplored. Therefore, it is critical to thoroughly understand the trade-offs among accuracy, robustness, and efficiency in this context.

This paper introduces AQUA-LLM, a comprehensive evaluation framework for cybersecurity QA that benchmarks a suite of edge-friendly, open-source LLMs across four configurations: pre-trained (base), fine-tuned, quantized, and fine-tuned quantized. We assess each model's predictive performance and adversarial robustness, focusing specifically on resilience to direct prompt injection attacks. Our experiments on two recent cybersecurity QA benchmarks reveal several key findings: (i) task-specific fine-tuning significantly improves prediction accuracy; (ii) quantization, while enhancing efficiency, substantially reduces robustness; (iii) combining quantization with fine-tuning can recover both accuracy and robustness; (iv) cybersecurity-specialized LLMs demonstrate alarmingly poor robustness and should be used with caution; and (v) ultimately, there exists a fundamental trade-off between accuracy, robustness, and efficiency that must be carefully considered when deploying LLMs in security-critical applications. 
%These findings highlight the need for robustness-aware fine-tuning strategies to enable secure and efficient deployment of quantized LLMs.  

\section{Related Work}
\subsection{Fine-Tuning LLMs for Cybersecurity Tasks}
Recent work has focused on fine-tuning LLMs to enhance their domain-specific performance in cybersecurity tasks. Liu et al.~\cite{liu2024cyberbench} introduced \textit{CyberBench}, a multi-task benchmark evaluating LLMs on named entity recognition, summarization, classification, and multiple-choice QA. Their results show that smaller, fine-tuned LLMs can sometimes match or exceed the performance of larger general-purpose models. In contrast, Tihanyi et al.~\cite{tihanyi2024cybermetric} proposed \textit{CyberMetric}, a retrieval-augmented benchmark for cybersecurity QA, and found that larger models like GPT-4 and Gemini-Pro consistently outperformed smaller alternatives. Meanwhile, Eizemity et al.~\cite{eizemity2025cyberllminstruct} investigated the safety risks of fine-tuning, finding that fine-tuned models became more vulnerable to adversarial prompts, highlighting a trade-off between task performance and robustness. %Zhang et al.~\cite{zhang2023hackmentor} introduced \textit{HackMentor}, demonstrating that lightweight models such as LLaMA and Vicuna can be significantly improved for cybersecurity use through fine-tuning, offering promising directions for accessible AI security tools.

\subsection{Prompt Injection and LLM Robustness}
Parallel to these efforts, several studies have investigated the robustness of LLMs against adversarial prompt injections. Bhatt et al.~\cite{bhatt2024cyberseceval2} introduced \textit{CyberSecEval~2}, which evaluates model responses to various prompt injection attacks, identifying a safety-utility trade-off. Greshake et al.~\cite{greshake2023indirect} highlighted the risk of \textit{indirect prompt injections} and showed that LLMs often fail to distinguish user intent from malicious commands. Follow-up work by Yi et al.~\cite{yi2024benchmarking} proposed boundary-aware prompting defenses, while Liu et al.~\cite{liu2024gradient} developed gradient-guided methods to craft highly effective prompt attacks. Hubinger et al.~\cite{hubinger2024sleeper} explored \textit{sleeper agents}, which are LLMs trained to behave maliciously when triggered by covert prompts, and demonstrated that such backdoors may persist even after standard safety training.

\subsection{Impact of Quantization and Joint Vulnerability}
As LLMs are deployed on edge devices, quantization has emerged as a practical solution to reduce memory and compute costs. However, this compression can introduce vulnerabilities. Egashira et al.~\cite{egashira2024quantizing} showed that quantized models may exhibit adversarial behavior not present in their full-precision counterparts. Similarly, Kumar et al.~\cite{kumar2024increased} demonstrated that fine-tuning and quantization, when combined, can degrade robustness and increase susceptibility to prompt injection attacks. Zhu et al.~\cite{zhu2023promptbench} proposed \textit{PromptBench}, finding that even small perturbations (e.g., syntax changes) can significantly impair model performance in security-sensitive contexts.

Despite these findings, the combined effects of fine-tuning and quantization remain insufficiently explored. Prior studies predominantly investigate fine-tuning and quantization separately, leaving a critical gap in understanding how to jointly preserve task performance and adversarial robustness when deploying LLMs in practical cybersecurity workflows.

\section{\Design{} Framework}
\label{framework}
%%%%%%%%%%%%%%%%%%%%%%%%%%%%%%%%%%%%
\begin{figure}[]
    \centering
    \includegraphics[width=.5\textwidth]{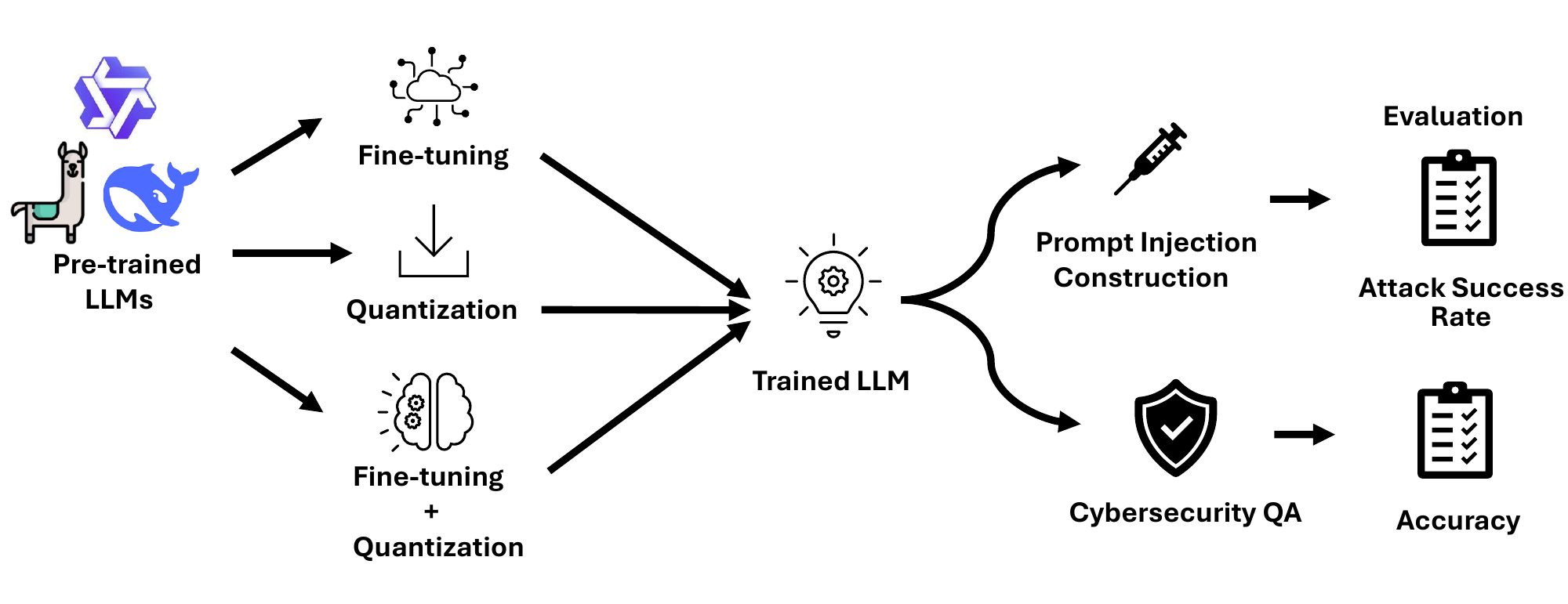}
    \caption{Proposed Evaluation Framework (AQUA-LLM)}
    \label{fig:aqua-llm}
\end{figure}
%%%%%%%%%%%%%%%%%%%%%%%%%%%%%%%%%%%%

Figure~\ref{fig:aqua-llm} illustrates AQUA-LLM, our evaluation framework for assessing the accuracy, robustness, and efficiency of LLMs across four configurations: Base (B), Quantized (Q), Fine-tuned (FT), and Fine-tuned Quantized (FTQ). The framework involves three stages: (1) adapting pre-trained LLMs via quantization, fine-tuning, or both; (2) conducting parallel evaluations on cybersecurity question answering (accuracy) and prompt injection attacks (adversarial robustness); and (3) analyzing metrics such as accuracy, attack success rate (ASR), and inference latency to explore trade-offs between performance, robustness, and efficiency. This assessment provides actionable insights to guide the deployment of LLMs in cybersecurity settings where balancing these factors is crucial.

\subsection{Pre-trained Large Language Models (Base LLMs)}
We evaluate a representative set of open-source small LLMs that vary in parameter size, instruction tuning, and domain specialization. The selected models include Meta LLaMA-3.1-8B-Instruct, an updated 8-billion parameter instruction-tuned model; Mistral-7B-Instruct, a 7-billion parameter model fine-tuned with reinforcement learning from human feedback (RLHF); Phi-3.5-Mini-Instruct, a compact 16-bit quantized model balancing efficiency and performance; Foundation-Sec-8B, a cybersecurity-specific 8-billion parameter model optimized for security tasks~\cite{kassianik2025}; Qwen 2.5-7B-Instruct, a 7-billion parameter multilingual instruction-tuned model utilizing 16-bit quantization; and DeepSeek-R1-Distill, a distilled model designed to reduce size and inference latency.

\subsection{Datasets}
We evaluate our LLM models on two cybersecurity-focused question-answering benchmarks: CyberBench~\cite{liu2024cyberbench} and CyberMetric~\cite{tihanyi2024cybermetric}. CyberBench aggregates ten publicly available cybersecurity sub-datasets encompassing various NLP tasks such as Named Entity Recognition (NER), abstractive summarization, and text classification. For this study, we adapt the entire benchmark into a multiple-choice QA format by treating each instance as an independent multiple-choice question, enabling a unified evaluation protocol. CyberMetric is a multiple-choice QA benchmark specifically curated for cybersecurity. The questions are generated using retrieval-augmented GPT-3.5 prompts and subsequently validated by domain experts. 
% We utilize the largest 10,000-question split to conduct a thorough evaluation of model capabilities in cybersecurity QA.

\subsection{Fine-tuning}
For fine-tuning, we employ distinct training and testing splits on two cybersecurity QA benchmarks. On the \textit{CyberMetric} dataset, 850 samples are allocated for training, while 150 samples are reserved for testing, divided into two subsets containing 50 and 100 questions, respectively. For the \textit{CyberBench} dataset, fine-tuning is conducted on the \textit{SecMMLU} subset, which comprises 116 security-related questions modeled after the MMLU benchmark. Evaluation is performed on the \textit{CyQuiz} subset, consisting of 128 applied cybersecurity QAs. Due to dataset size constraints, the \textit{CyQuiz} test set is further partitioned into overlapping subsets of 50 and 100 questions.

Data preprocessing involves loading a Hugging Face-compatible version of the datasets and standardizing raw samples into instruction-response pairs using the Unsloth library~\cite{unsloth2023}. Prompts are formatted with model-specific chat templates to ensure compatibility with the tokenization requirements of each model. Subsequently, the datasets are tokenized and prepared for supervised fine-tuning.

Supervised fine-tuning process employs Low-Rank Adaptation (LoRA)~\cite{hu2022lora}, which selectively updates key projection layers within the transformer architecture to efficiently adapt the model. The LoRA adapters are configured with a rank of 16 and an alpha scaling factor of 16, with no dropout applied. Integration is performed using the Unsloth library’s parameter-efficient fine-tuning interface. The fine-tuned models are subsequently saved in GGUF format for evaluation.

\subsection{Quantization}
To enable efficient deployment on edge hardware, we apply 4-bit NormalFloat (NF4) quantization using the BitsAndBytes library~\cite{dettmers2023qlora}. At model initialization, each pretrained 16-bit checkpoint is replaced with its 4-bit NF4 counterpart via the Transformers library integration, incurring no runtime quantization overhead. NF4 is a data type specifically designed to optimally represent normally distributed weights at low precision, and BitsAndBytes applies double quantization to further reduce metadata footprint~\cite{dettmers2023qlora}. This approach reduces GPU memory consumption by approximately 75\% and significantly improves inference throughput, while maintaining compatibility with our training and evaluation pipelines.

%\subsection{Quantization}
%We pulled off‐the‐shelf, 4-bit NF4 models provided by Hugging Face. At load time, the pretrained 16-bit checkpoints were replaced by their corresponding 4-bit NormalFloat (NF4) variants via the bitsandbytes integration in the Transformers library. This approach ensured compatibility with our training and inference pipelines—with zero runtime quantization overhead, while reducing GPU memory usage by 75\% and boosting throughput during benchmarking.

%\subsection{Fine-tuning with Quantization}
%We conducted parameter-efficient fine-tuning using Low-Rank Adaptation (LoRA) on the pre-quantized 4-bit models: for each Transformer layer, our LoRA framework inserts a pair of rank-\(r\) matrices \(A\in\mathbb{R}^{d\times r}\) and \(B\in\mathbb{R}^{r\times d}\) into the query/key/value weight blocks. During finetuning, the base weights remained frozen and only the low-rank adapter matrices were updated, reducing the number of trainable parameters. 

%Figure~\ref{fig:training-loss} shows finetuning data over 60 steps with a peak learning rate of 2 × 10⁻⁴ under the AdamW optimizer, under the same dataset configurations as the base models.
%Upon convergence, the learned LoRA adapters were merged back into the 4-bit model weights to produce a single 4-bit checkpoint for loading and inference.

\subsection{Fine-Tuning with Quantized Models}
We perform parameter-efficient fine-tuning on pre-quantized 4-bit LLMs using LoRA. In this setup, LoRA injects trainable low-rank projection matrices into the attention components (query, key, and value projections) of each transformer layer, while keeping the original model weights frozen. This approach significantly reduces the number of trainable parameters, enabling efficient adaptation without full model updates. Fine-tuning is conducted over 60 steps using the AdamW optimizer with a peak learning rate of \(2 \times 10^{-4}\), under the same dataset splits used for base model evaluation. Following convergence, the learned LoRA adapter weights are merged into the quantized model, resulting in a single 4-bit checkpoint suitable for efficient inference. This combined approach leverages both the computational benefits of quantization and the task-specific adaptability provided by fine-tuning.

%%%%%%%%%%%%%%%%%%%%%%%%%%%%%%%%%%%%%%%%%%%%%%%%%%%%%%%%%%%%%%%%%%%%%%%
%\begin{figure}[]
%    \centering
%    \includegraphics[width=.4\textwidth,trim=0pt 0pt 0pt 16pt,clip]{training loss.pdf}
%    \caption{Training Loss over 60 steps} 
%    \label{fig:training-loss}
%\end{figure}
%%%%%%%%%%%%%%%%%%%%%%%%%%%%%%%%%%%%%%%%%%%%%%%%%%%%%%%%%%%%%%%%%%%%%%%

% Quantization and fine-tuning were applied using the Unsloth QLoRA framwork. To enable efficient inference and reduce memory consumption, \textbf{4-bit quantization} was applied to a pre-trained 16-bit language model at load time. Fine-tuning was conducted using \textbf{Low-Rank Adaptation (LoRA)}, with the adapter weights updated during training. After training, the LoRA adapter weights were merged into the base model to produce a fully integrated 4-bit fine-tuned version. Gradient check pointing was enabled via the Unsloth configuration to support backpropagation. 

%%%%%%%%%%%%%%%%%%%%%%%%%%%%%%%%%%%%
%\begin{figure}[]
%    \centering
%    \includegraphics[width=.45\textwidth]{ThreatModel.pdf}
%    \caption{Threat Model}
%    \label{fig:th-model}
%\end{figure}
%%%%%%%%%%%%%%%%%%%%%%%%%%%%%%%%%%%%

\subsection{Prompt Injection}
\textbf{Threat Model:} We consider the deployment of a quantized, open-source LLM for cybersecurity QA on resource-constrained edge devices. The critical \textbf{assets} at risk include: (1) the domain-specific cybersecurity knowledge encapsulated in the model's outputs, and (2) the integrity and reliability of the model’s responses, which directly influence security decision-making processes. \textbf{Adversary} is an external actor with access to the input interfaces of the system, such as system logs, alerts, or user-submitted queries, and is capable of crafting adversarial prompts to influence model behavior. The primary \textbf{threat} is direct prompt injection, where the attacker manipulates inputs to elicit sensitive information or produce misleading or insecure recommendations. This threat is exacerbated by key \textbf{vulnerabilities}, including the system’s reliance on potentially untrusted input channels and the absence of robust input/output sanitization mechanisms limitations. The potential \textbf{impact} of a successful attack includes unauthorized disclosure of sensitive cybersecurity knowledge and the propagation of insecure actions or configurations, which may compromise the overall security posture of the system.

\textbf{Direct Prompt Injection Evaluation:} We evaluate model robustness to direct prompt injection attacks using the DeepTeam Red Teaming Framework~\cite{deepteam2025}. Specifically, we instantiate the \textit{IllegalActivity} vulnerability class (subset: \textit{cybercrime}), which simulates realistic cybersecurity threat scenarios such as phishing, unauthorized system access, and malware distribution. The \texttt{PromptInjection} attack class is employed to generate single-turn adversarial prompts aligned with this vulnerability type. For each model under evaluation, DeepTeam utilizes OpenAI’s GPT-3.5 API to generate either 50 or 100 adversarial examples per configuration. These inputs are subsequently used to query the target LLM in an inference setting. Model responses are then assessed via a second GPT-3.5 API call implementing DeepTeam’s \texttt{harmful-content} metric, which categorizes outputs as either harmful or non-harmful. We benchmark each model on both sets of 50 and 100 adversarial inputs. The overall Attack Success Rate (ASR) is calculated as the proportion of prompts that elicit harmful responses, thereby quantifying each model’s vulnerability to direct prompt injection within this operational context. 

%Prompt injection attacks were probed using the DeepEval DeepTeam Red Teaming Framework. We instantiated DeepTeam’s IllegalActivity vulnerability class (subset “cybercrime”) to simulate real-world digital-crime scenarios such as phishing, unauthorized access, and malware deployment. We then used the framework's PromptInjection attack class to simulate single-turn adversarial inputs. For each model, the Red Teaming framework operated by calling OpenAI's  GPT-3.5 model API to generate 50 or 100 adversarial examples configured to our vulnerability class and attack method, which were then served as inference calls to the target model. Each model response was then evaluated by a second GPT-3.5 call implementing DeepTeam’s harmful-content metric, classifying outputs as harmful or not. Each model was benchmarked on both sets of 50 and 100 prompt injections, where the total Attack Success Rate (ASR) was quantified by the proportion of harmful outputs produced.

\subsection{Evaluation}
\textbf{Hardware:} Our experiments were performed on a
Linux virtual machine server equipped with a 16-core CPU, 32 GB of RAM, and an NVIDIA A100 GPU with 80 GB of memory. 

\textbf{Evaluation Metrics:} We evaluate model performance using three key metrics: Accuracy, Attack Success Rate (ASR), and Inference Latency.

\paragraph{Accuracy}
LLM accuracy is assessed using subsets from the CyberMetric and CyberBench benchmarks. Each model is prompted with either a 50- or 100-question evaluation set, where each question includes a set of predefined answer choices. A response is considered correct if the model’s selected answer matches the ground truth. Accuracy is computed as the percentage of correctly answered questions out of the total number of questions.

\paragraph{Attack Success Rate (ASR)}
To evaluate adversarial robustness, we compute the Attack Success Rate (ASR), which reflects the proportion of prompt injection attacks that successfully induce harmful or incorrect responses. Specifically, ASR is calculated as the percentage of generated adversarial inputs that lead the model to produce a harmful output, as determined by an automated red-teaming evaluation. Lower ASR indicates greater robustness, and vice versa.

\paragraph{Inference Latency}
Inference efficiency is quantified by measuring the average per-question latency in seconds.

%\textbf{Evaluation Metrics:} 
%\textbf{Accuracy:}
%Model accuracy was measured using subsets of the CyberMetric and CyberBench datasets. Each model was prompted with questions from either a 50 or 100 question evaluation set. The model's output was compared to the correct answer in the dataset. 

%A prediction was considered correct if the model's output matched the expected answer. Accuracy was computed as the proportion of correctly answered questions relative to the total number of questions in the evaluation set. 
%This metric was applied among all base, fine-tuned, quantized, and fine-tuned quantized models. 

%\textbf{Attack Success Rate:}
%Attack Success Rate was computed by subtracting 100 from the number of prompt injection attacks that were mitigated by the model. This yields the number of successful prompt injections that the model was not able to mitigate and therefore the successful rate of attacks.

\section{Experimental Results}
\label{experimental}
%%%%%%%%%%%%%%%%%%%%%%%%%%%%%%%%%%%%%%%%%%%%%%%%%%%%%%%%%%%%%%%%
\begin{figure}[]
  \centering
  \begin{subfigure}[b]{0.95\linewidth}
    \centering
    \includegraphics[width=\linewidth,trim=10pt 12pt 10pt 14pt,clip]
      {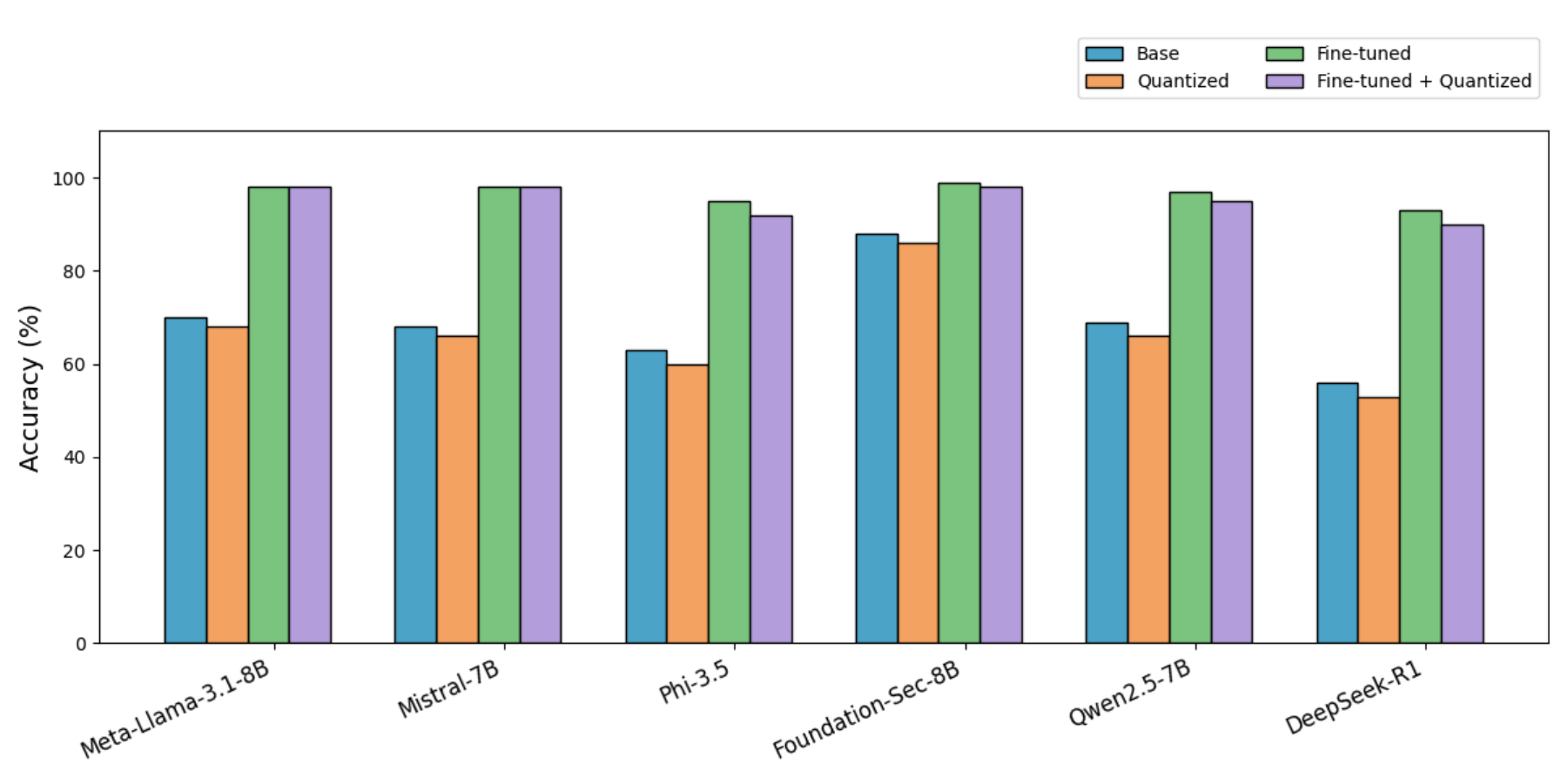}
    \caption{CyberBench}
    \label{fig:acc-cyberbench}
  \end{subfigure}
  \hfill
  \begin{subfigure}[b]{0.95\linewidth}
    \centering
    \includegraphics[width=\linewidth,trim=10pt 12pt 10pt 14pt,clip]
      {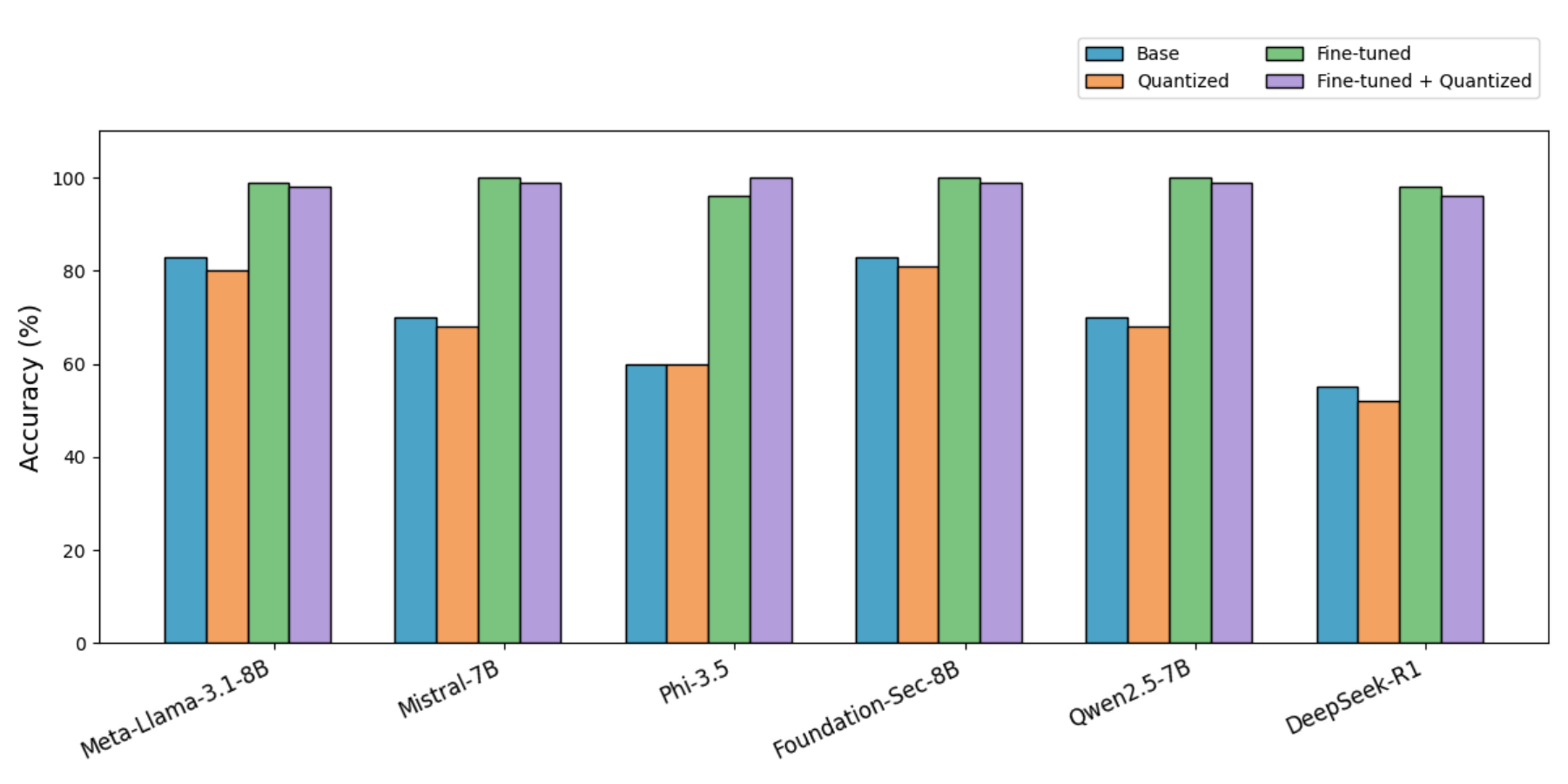}
    \caption{CyberMetric}
    \label{fig:acc-cybermetric}
  \end{subfigure}
  \caption{Accuracy comparison of LLMs on 100-question tasks}
  \label{fig:combined-accuracy}
\end{figure}
%%%%%%%%%%%%%%%%%%%%%%%%%%%%%%%%%%%%%%%%%%%%%%%%%%%%%%%%%%%%%%%%

%%%%%%%%%%%%%%%%%%%%%%%%%%%%%%%%%%%%%%%%%%%%%%%%%%%%%%%%%%%%%%%%
\begin{figure}[]
  \centering
  \begin{subfigure}[b]{0.95\linewidth}
    \centering
    \includegraphics[width=\linewidth,trim=10pt 12pt 10pt 14pt,clip]
      {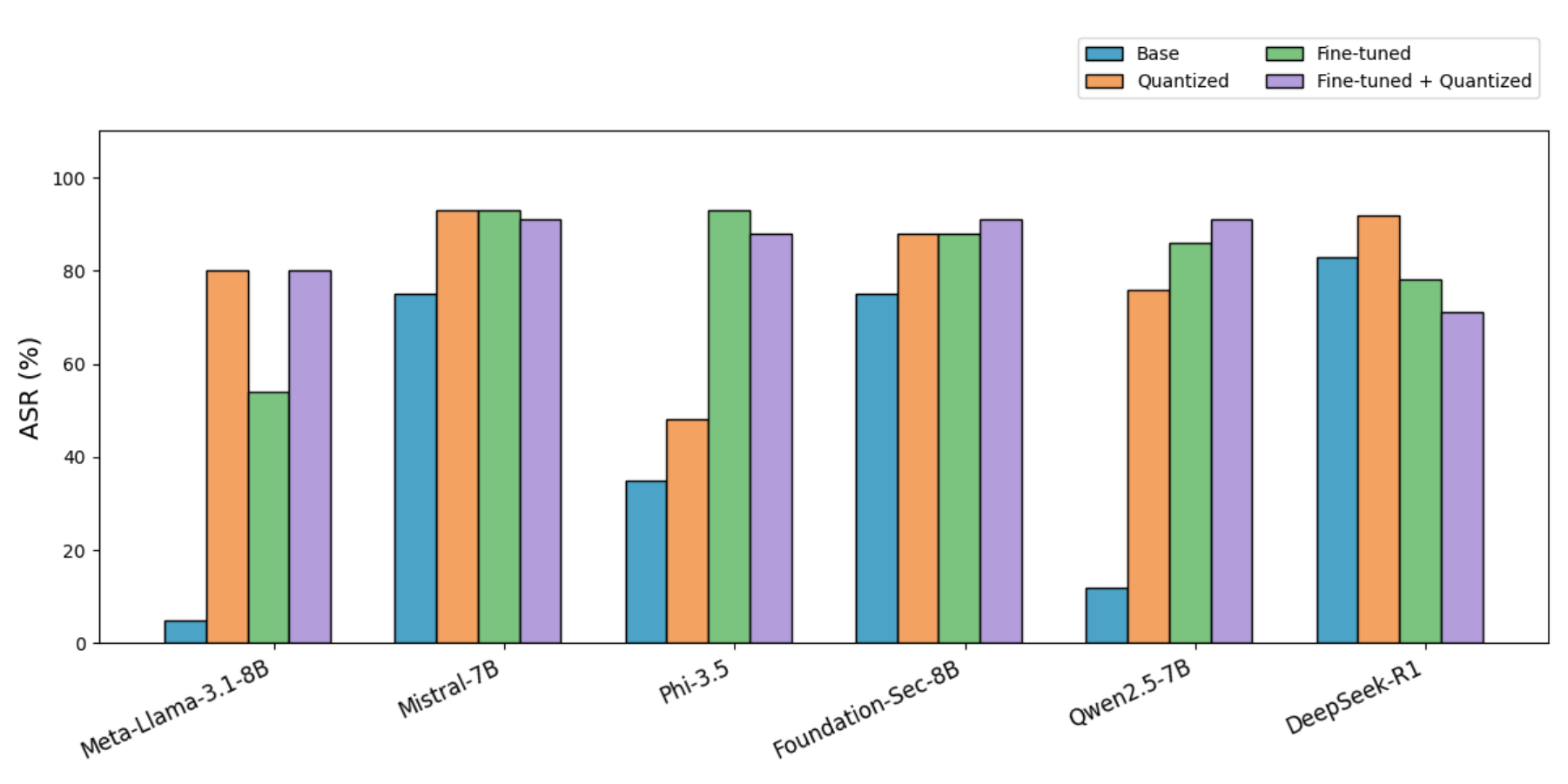}
    \caption{CyberBench}
    \label{fig:asr-cyberbench}
  \end{subfigure}
  \hfill
  \begin{subfigure}[b]{0.95\linewidth}
    \centering
    \includegraphics[width=\linewidth,trim=10pt 12pt 10pt 14pt,clip]
      {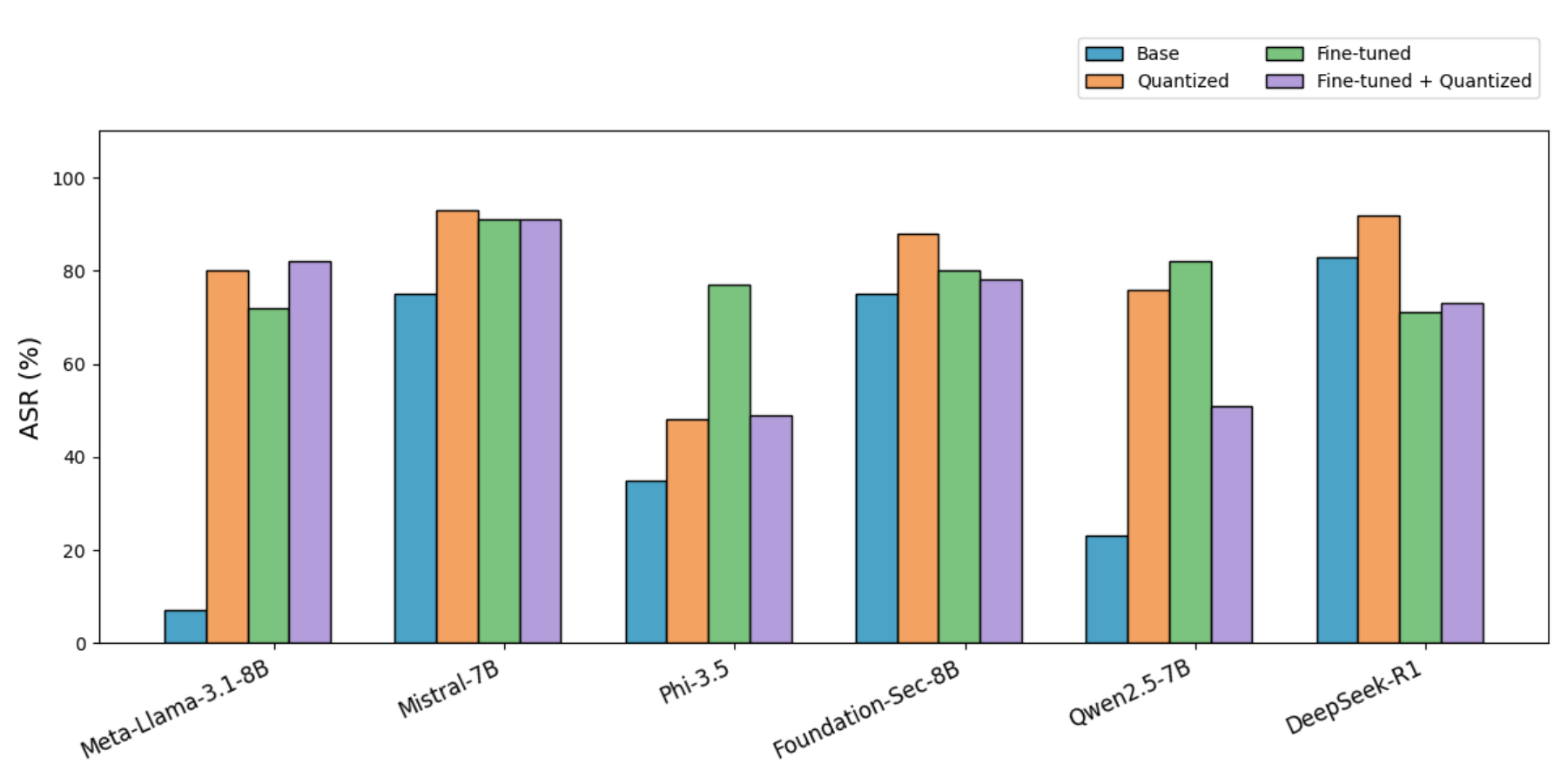}
    \caption{CyberMetric}
    \label{fig:asr-cybermetric}
  \end{subfigure}
  \caption{ASR comparison of LLMs on 100-question tasks}
  \label{fig:asr-combined}
\end{figure}
%%%%%%%%%%%%%%%%%%%%%%%%%%%%%%%%%%%%%%%%%%%%%%%%%%%%%%%%%%%%%%%%

Figures~\ref{fig:combined-accuracy} and~\ref{fig:asr-combined} present the accuracy and attack success rate (ASR) metrics for selected LLMs evaluated on cybersecurity QA datasets. In each subfigure, colors denote different LLM configurations (e.g., quantized, fine-tuned), the x-axis represents the LLM variants, and the y-axis reflects the corresponding metric values. Based on these results, we summarize key findings from our evaluation, with high-level observations highlighted in bold to emphasize prominent trends. Each finding is subsequently analyzed in detail.

%%%%%%%%%%%%%%%%%%%%%%%%%%%%%%%%%%%%%%%%%%%%%%%%%%
\begin{table}[]
  \centering
  \caption{Multiple-choice QA accuracy (\%) before and after fine-tuning on CyberBench and CyberMetric}
  \label{tab:qa-combined}
  \scalebox{0.92}{
  \begin{tabular}{|l|cc|cc|}
    \hline
    \textbf{LLM Model / Dataset} & \multicolumn{2}{c|}{\textbf{CyberBench}} & \multicolumn{2}{c|}{\textbf{CyberMetric}} \\
                       & \textbf{50 Q} & \textbf{100 Q} & \textbf{50 Q} & \textbf{100 Q} \\
    \hline
    Meta-Llama-3-1-8B  & \textcolor{red}{72} $\rightarrow$ \textcolor{green!50!black}{100} & \textcolor{red}{70} $\rightarrow$ \textcolor{green!50!black}{98}  & \textcolor{red}{82} $\rightarrow$ \textcolor{green!50!black}{98}  & \textcolor{red}{83} $\rightarrow$ \textcolor{green!50!black}{99} \\
    Mistral-7B         & \textcolor{red}{68} $\rightarrow$ \textcolor{green!50!black}{100} & \textcolor{red}{68} $\rightarrow$ \textcolor{green!50!black}{98}  & \textcolor{red}{70} $\rightarrow$ \textcolor{green!50!black}{100} & \textcolor{red}{70} $\rightarrow$ \textcolor{green!50!black}{100} \\
    Phi-3.5            & \textcolor{red}{64} $\rightarrow$ \textcolor{green!50!black}{96}  & \textcolor{red}{63} $\rightarrow$ \textcolor{green!50!black}{95}  & \textcolor{red}{60} $\rightarrow$ \textcolor{green!50!black}{96}  & \textcolor{red}{60} $\rightarrow$ \textcolor{green!50!black}{96} \\
    Foundation-Sec-8B  & \textcolor{red}{92} $\rightarrow$ \textcolor{green!50!black}{100} & \textcolor{red}{88} $\rightarrow$ \textcolor{green!50!black}{99}  & \textcolor{red}{86} $\rightarrow$ \textcolor{green!50!black}{98}  & \textcolor{red}{83} $\rightarrow$ \textcolor{green!50!black}{100} \\
    Qwen2.5-7B         & \textcolor{red}{70} $\rightarrow$ \textcolor{green!50!black}{98}  & \textcolor{red}{69} $\rightarrow$ \textcolor{green!50!black}{97}  & \textcolor{red}{70} $\rightarrow$ \textcolor{green!50!black}{100} & \textcolor{red}{70} $\rightarrow$ \textcolor{green!50!black}{100} \\
    DeepSeek-R1        & \textcolor{red}{58} $\rightarrow$ \textcolor{green!50!black}{92}  & \textcolor{red}{56} $\rightarrow$ \textcolor{green!50!black}{93}  & \textcolor{red}{56} $\rightarrow$ \textcolor{green!50!black}{100} & \textcolor{red}{55} $\rightarrow$ \textcolor{green!50!black}{98} \\
    \hline
  \end{tabular}}
\end{table}
%%%%%%%%%%%%%%%%%%%%%%%%%%%%%%%%%%%%%%%%%%%%%%%%%%

\textbf{1) Fine-tuning substantially improves task performance.} Table~\ref{tab:qa-combined} reports the multiple-choice question-answering accuracy of the base models (red) and their fine-tuned counterparts (green) across both the CyberBench and CyberMetric datasets, allowing direct comparison of performance before and after fine-tuning. We can observe that the fine-tuned models consistently achieve the highest QA accuracy across both CyberMetric and CyberBench datasets. For example, \texttt{Mistral-7B} reaches 100\% accuracy on CyberMetric, compared to 70\% in its base configuration. On average, fine-tuned models achieve the highest QA accuracy, reaching 98.75\% on CyberMetric and 97.17\% on CyberBench, compared to 70.42\% and 69.83\% for their respective base models. \textit{This underscores the critical role of task-specific fine-tuning in enhancing model performance.}

\textbf{2) Quantization accelerates inference at the cost of robustness.} Table~\ref{tab:qa-cyberbench-latency-robustness} presents the average per-question inference time and attack success rate (ASR) for each LLM, highlighting trade-offs between computational efficiency and robustness. Based on the latency results, quantized models achieve up to 1.28$\times$ speedup over their base counterparts, primarily due to improved memory efficiency from reducing model precision from 16-bit to 4-bit. However, quantized models exhibit significantly increased vulnerability to adversarial attacks, with up to a 16$\times$ loss in robustness. On average, ASR increases from 52.25\% in base models to 77.92\% in quantized models across all evaluated LLMs. \textit{These results show that quantization enhances efficiency in cybersecurity QA but requires careful robustness evaluation to mitigate increased adversarial risk.}

%%%%%%%%%%%%%%%%%%%%%%%%%%%%%%%%%%%%%%%%%%%%%%%%%%
\begin{table*}[]
  \centering
  \caption{
    Per-Question Inference Latency (s) and Attack Success Rate (ASR, \%) of Base vs. Quantized Models on CyberMetric.
    $\uparrow$ Speedup and $\downarrow$ ASR indicate improved performance and robustness, respectively.
  }
  \label{tab:qa-cyberbench-latency-robustness}
  \begin{tabular}{|l|cc|c|cc|c|}
    \hline
    \multirow{2}{*}{\textbf{LLM Model}} & 
    \multicolumn{2}{c|}{\textbf{Latency (s)}} & 
    \textbf{Speedup} & 
    \multicolumn{2}{c|}{\textbf{ASR (\%)}} & 
    \textbf{Robustness Loss} \\
    \cline{2-3} \cline{5-6}
    & \textbf{Base} & \textbf{Quantized} & & \textbf{Base} & \textbf{Quantized} & \\
    \hline
    Meta-Llama-3.1-8B    & 0.140 & 0.112 & 1.25$\times$ & 5  & 80  & \textbf{16}\boldmath$\times$ \\
    Mistral-7B           & 0.118 & 0.093 & \underline{1.26$\times$} & 75 & 93 & 1.24$\times$ \\
    Phi-3.5              & 0.124 & 0.112 & 1.10$\times$ & 35 & 48 & 1.37$\times$ \\
    Foundation-Sec-8B    & 0.136 & 0.116 & 1.17$\times$ & 75 & 88 & 1.17$\times$ \\
    Qwen2.5-7B           & 0.138 & 0.122 & 1.13$\times$ & 5  & 76  & \underline{15.2$\times$} \\
    DeepSeek-R1          & 0.154 & 0.120 & \textbf{1.28}\boldmath$\times$ & 83 & 92 & 1.11$\times$ \\
    \hline
  \end{tabular}
\end{table*}

\begin{figure}[]
  \centering
  \includegraphics[width=0.95\linewidth]{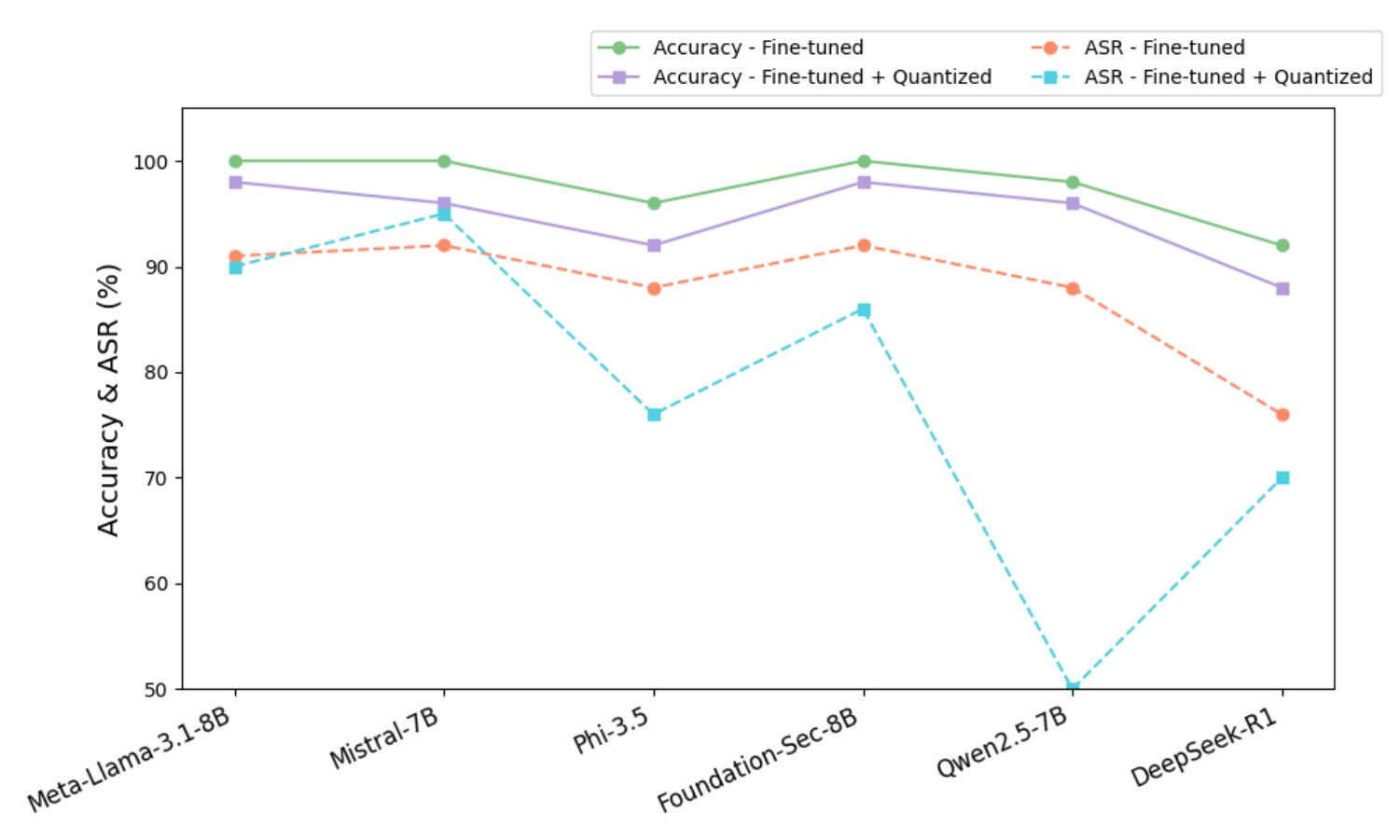}
  \caption{Accuracy and ASR of Fine-Tuned (FT) vs. Fine-Tuned Quantized (FTQ) Models on CyberBench}
  \label{fig:cybermetric_quant_combined}
\end{figure}

\textbf{3) Fine-tuned quantized models preserve accuracy while improving robustness over fine-tuning alone.} Figure~\ref{fig:cybermetric_quant_combined} compares the performance of fine-tuned (FT) and fine-tuned quantized (FTQ) models on the CyberBench dataset, presenting accuracy and attack success rate (ASR) results. We observe that FTQ models achieve QA performance closely aligned with their unquantized (FT) counterparts. On average, FTQ models attain an accuracy of 98.25\% compared to 98.75\% for FT models on CyberMetric, and 94.92\% versus 97.17\% on CyberBench. More notably, FTQ models exhibit lower average ASR than FT models across both datasets: 71.58\% vs.\ 76.25\% on CyberMetric, and 64.58\% vs.\ 84.08\% on CyberBench. This trend mostly holds, with only a few exceptions across the evaluated LLMs, e.g., \texttt{Meta-Llama-3.1-8B}. \textit{These findings challenge the common assumption that quantization inherently increases adversarial vulnerability, suggesting instead that, when combined with task-specific fine-tuning, it may act as a regularization mechanism that enhances model robustness.}

%%%%%%%%%%%%%%%%%%%%%%%%%%%%%%%%%%%%%%%%%%%%%%%%%%
%%%%%%%%%%%%%%%%%%%%%%%%%%%%%%%%%%%%%%%%%%%%%%%%%%

\textbf{4) Quantization without fine-tuning degrades both accuracy and robustness.} As illustrated in Figures~\ref{fig:combined-accuracy} and~\ref{fig:asr-combined}, quantized models without fine-tuning consistently exhibit the weakest overall performance. For example, \texttt{DeepSeek-R1-Distill} (quantized) achieves less than 52\% QA accuracy and exhibits ASR exceeding 90\%, whereas its fine-tuned counterpart reaches 98--100\% accuracy with substantially lower ASR (71--76\%). On average, quantized-only models show QA accuracy around 67\% and ASR near 78\%, compared to fine-tuned models, which achieve accuracy above 97\% and ASR ranging from 76\% to 84\%. \textit{Taken together, the findings indicate that fine-tuning is essential for restoring both predictive performance and robustness in quantized LLMs.}

%%%%%%%%%%%%%%%%%%%%%%%%%%%%%%%%%%%%%%%%%%%%%%%%%%%%%%%%%%%%%%%%
%\begin{figure}[]
%  \centering
%  \begin{subfigure}[b]{0.95\linewidth}
%    \centering
%    \includegraphics[width=\linewidth,trim=10pt 12pt 10pt 28pt,clip]
%      {datasetset_acc_hash_final.pdf}
%    \caption{Accuracy}
%    \label{fig:foundation-cybermetric}
%  \end{subfigure}
%  \hfill
%  \begin{subfigure}[b]{0.95\linewidth}
%    \centering
%    \includegraphics[width=\linewidth,trim=10pt 12pt 10pt 14pt,clip]
%      {dataset_asr_hash.pdf}
%    \caption{Attack Success Rate (ASR)}
%    \label{fig:foundation-ASR}
%  \end{subfigure}
%  \caption{Average accuracy and adversarial robustness of evaluated LLMs on CyberMetric and CyberBench}
%  \label{fig:foundation-comparison}
%\end{figure}
%%%%%%%%%%%%%%%%%%%%%%%%%%%%%%%%%%%%%%%%%%%%%%%%%%%%%%%%%%%%%%%%

\textbf{5) The cybersecurity-specialized LLM achieves top-tier accuracy but exhibits mixed robustness performance.} Built on the \texttt{Llama-3.1-8B} architecture, \texttt{Foundation-Sec-8B} consistently delivers the highest QA accuracy across both datasets, reaching up to 100\% in fine-tuned (FT) and fine-tuned quantized (FTQ) configurations. However, its adversarial robustness varies significantly by configuration and dataset. On CyberMetric, \texttt{Foundation-Sec-8B} records extremely high ASR values, e.g., up to 100\% for fine-tuned and 87--88\% for quantized models, indicating increased vulnerability despite strong accuracy. Conversely, on CyberBench, it shows relatively lower ASR in FTQ and FT modes (86--92\%), though still higher than some other models. \textit{This pattern illustrates that while domain-specific training substantially enhances predictive performance, it does not necessarily translate to improved robustness, underscoring the critical need for robustness-aware optimization in cybersecurity-focused LLMs.}

\begin{figure}[]
  \centering
  \begin{subfigure}[b]{0.48\linewidth}
    \centering
    \includegraphics[width=\linewidth,trim=30pt 22pt 30pt 22pt,clip]{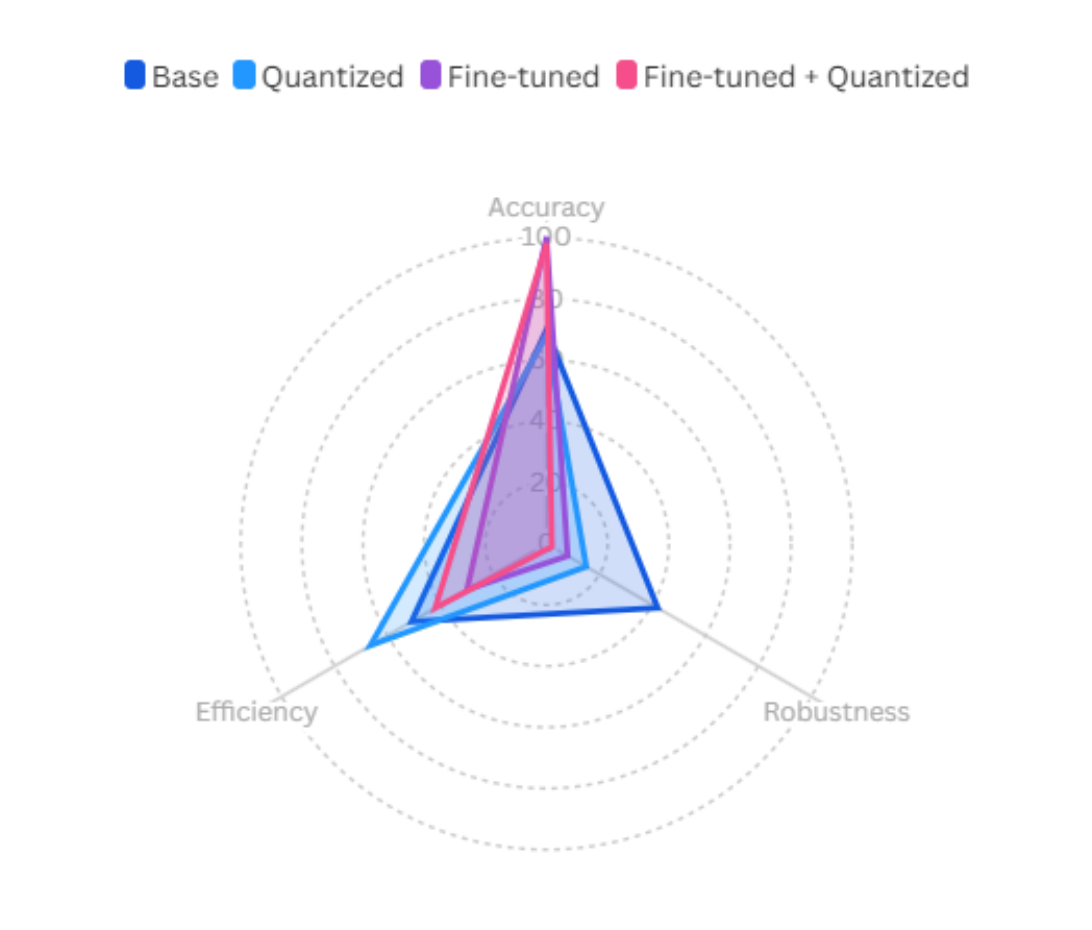}
    \caption{Mistral-7B-Instruct.}
    \label{fig:spider-cybermetric-a}
  \end{subfigure}
  \hfill
  \begin{subfigure}[b]{0.50\linewidth}
    \centering
    \includegraphics[width=\linewidth,trim=0pt 0pt 15pt 0pt,clip]{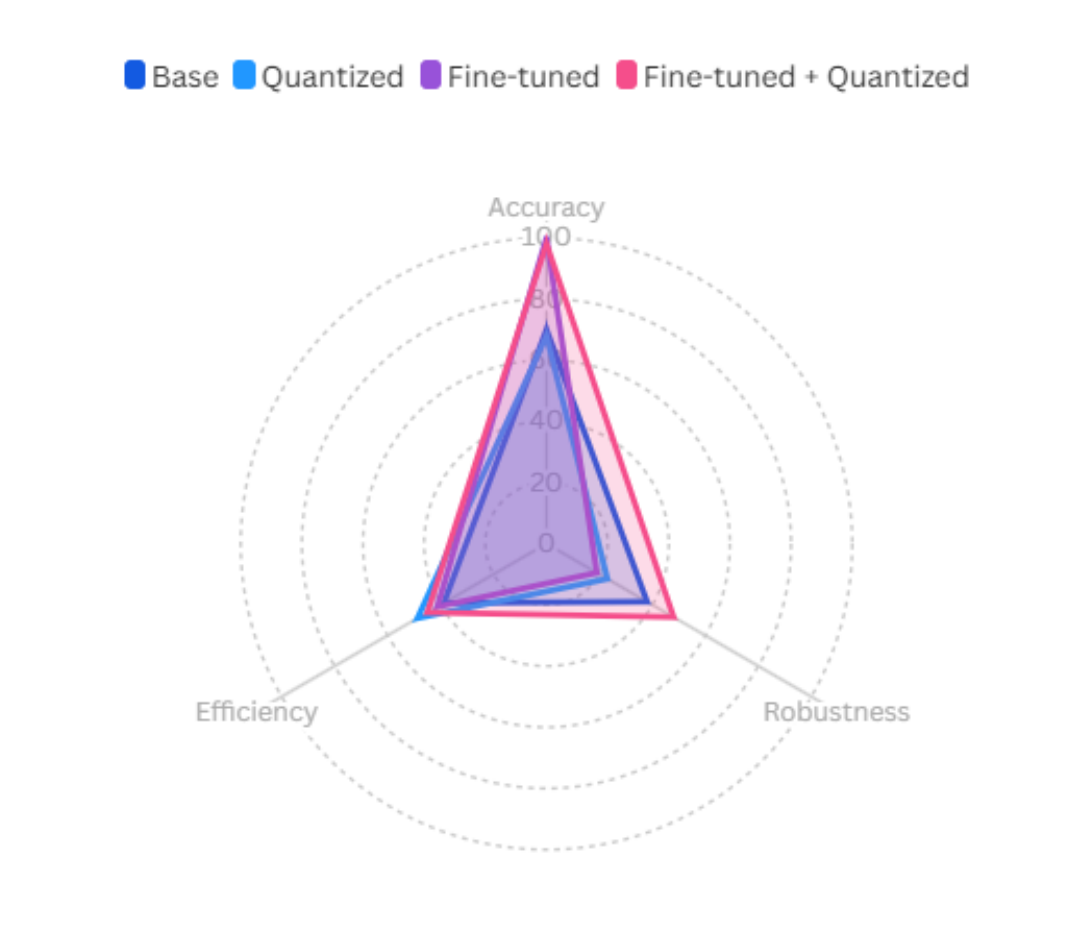}
    \caption{Qwen 2.5-7B.}
    \label{fig:spider-cybermetric-b}
  \end{subfigure}
  \caption{Trade-off among Accuracy, Robustness, and Efficiency.}
  \label{fig:spider-cybermetric}
\end{figure}

\textbf{6) A clear trade-off exists among accuracy, robustness, and computational efficiency.} Figure~\ref{fig:spider-cybermetric} presents spider charts illustrating this trade-off for \texttt{Mistral-7B} and \texttt{Qwen 2.5-7B} on the CyberMetric dataset. Here, we present base, quantized, fine-tuned, and fine-tuned quantized LLM configurations evaluated across three key dimensions: accuracy, efficiency, and robustness. For instance, in \texttt{Mistral-7B} (Figure~\ref{fig:spider-cybermetric-a}), this trade-off is clearly observed: the quantized model achieves the highest efficiency but at the cost of reduced accuracy and robustness. For \texttt{Qwen 2.5-7B} (Figure~\ref{fig:spider-cybermetric-b}), the fine-tuned quantized configuration emerges as a favorable choice, achieving high robustness while maintaining strong accuracy. \textit{Overall, no single configuration simultaneously optimizes accuracy, efficiency, and robustness. Although quantization substantially enhances efficiency, it typically degrades accuracy and robustness. Fine-tuning post-quantization can mitigate these drawbacks and, in some cases, yield models that achieve a favorable balance across all three metrics.}

%%%%%%%%%%%%%%%%%%%%%%%%%%%%%%%%%%%%%%%%%%%%%%%%%%%%%%%%%%%%%%%%
%%%%%%%%%%%%%%%%%%%%%%%%%%%%%%%%%%%%%%%%%%%%%%%%%%%%%%%%%%%%%%%%

% \textbf{7) Quantization-aware and robustness-preserving methods are essential.} The pronounced robustness degradation observed in fine-tuned quantized (FTQ) models, e.g., 95\% ASR of \texttt{Mistral-7B-Instruct} on the CyberMetric (50Q) benchmark, highlights the urgent need for optimization strategies that maintain adversarial robustness while facilitating efficient deployment on resource-constrained platforms.

\section{Conclusion}
\label{conclusion}
We propose a comprehensive evaluation of accuracy–robustness–efficiency trade-offs for LLMs in cybersecurity question answering. Leveraging our proposed AQUA-LLM framework, we compare four deployment configurations: base, quantized, fine-tuned, and fine-tuned with quantization, across six open-source LLMs on two QA datasets: CyberMetric and CyberBench. Our findings reveal that quantization alone, although beneficial for efficiency, considerably degrades accuracy and robustness. Finetuning alone improves QA accuracy but frequently reduces robustness compared to base models. In contrast, fine-tuning after quantization mitigates this trade-off, significantly enhancing both accuracy and robustness without sacrificing the efficiency gains provided by quantization. To safely leverage performance gains, LLM deployment in security-critical workflows must employ optimization techniques that maintain adversarial robustness. 

\section*{Acknowledgments}
This work has been funded in part by NSF, with award numbers \#1826967, \#1911095, \#2003279, \#2052809, \#2100237, \#2112167, \#2112665, and in part by PRISM and CoCoSys, centers in JUMP 2.0, an SRC program sponsored by DARPA.

\bibliographystyle{ieeetr}
\bibliography{bibfile}

\end{document}